\documentclass[prd,onecolumn,aps,nofootinbib,notitlepage,nobibnotes,superscriptaddress,preprintnumbers]{revtex4-1}

\usepackage[utf8]{inputenc}
\usepackage{color}%to be able to use colors
\usepackage{amsmath}%Recomended for mathematical documents (use of: eqref,...)
\usepackage{amsfonts}
\usepackage{graphicx}
\usepackage[dvipsnames]{xcolor}
\definecolor{refs}{RGB}{245,156,74}
 \usepackage[colorlinks=true,hyperfootnotes=true,citecolor=cyan]{hyperref}
 
\usepackage{dsfont} %Identity matrix

\newcommand{\dd}{{\rm d}}

\newcommand{\mS}{\mathcal{S}}

\newcommand{\be}{\begin{equation}}
\newcommand{\ee}{\end{equation}}
\newcommand{\bea}{\begin{eqnarray}}
\newcommand{\eea}{\end{eqnarray}}

\renewcommand{\bf}[1]{{\textbf{#1}}}

\newcommand{\mpl}{M_{\rm Pl}}
\newcommand{\pECG}{\beta}

\newcommand{\eN}{\mathrm{e}}
\newcommand{\eanis}{\epsilon_\sigma}

\begin{document}

\title{On the strong coupling of Einsteinian Cubic Gravity and its generalisations}

\author{Jose Beltr\'an Jim\'enez}
\email[]{jose.beltran@usal.es}
\affiliation{Departamento de F\'isica Fundamental and IUFFyM, Universidad de Salamanca, E-37008 Salamanca, Spain.}

\author{Alejandro Jim\'enez-Cano}
\email[]{alejandrojc@ugr.es}
\affiliation{Departamento de F\'isica Te\'orica y del Cosmos and CAFPE, Universidad de Granada, 18071 Granada, Spain.}

\begin{abstract}
In this note we discuss the strong coupling issues inherent to the defining requirement for the so-called Einsteinian Cubic Gravity and its quasi-topological generalisations.
\end{abstract}

\maketitle

\section{Introduction}

Einsteinian Cubic Gravity (ECG) is a higher curvature theory of gravity defined to possess the same linear spectrum as General Relativity (GR), i.e., a massless spin-2 field, around maximally symmetric spacetimes in arbitrary dimension \cite{Oliva:2010eb,Myers:2010ru,Bueno:2016xff}. As it is well-known, the only higher curvature theories that share the field content of GR around any background or, in other words, at the full non-linear order are provided by the Lovelock terms \cite{Lovelock1970,Lovelock1971}. Einsteinian Cubic Gravity falls outside this family of theories and, consequently, they contain additional degrees of freedom which in turn are associated to the higher than second order nature of the field equations\footnote{More precisely, one should say that these theories are neither any of the Lovelock Lagrangians nor are they related via a regular field redefinition to them. See \cite{Bueno:2019ltp} for an interesting discussion on the role of field redefinitions within the framework of generalised quasi-topological theories where it is shown that they formed a complete basis for the Lagrangians of the type $\mathcal{L}(g_{\mu\nu},R_{\mu\nu\rho\lambda})$.}. Among these additional dof's there will be ghostly modes associated to an Ostrogradski instability (see e.g. \cite{woodard2019theorem, Woodard:2006nt}). In an attempt to avoid this pathology, ECG selects a particular cubic polynomial of the Riemann tensor so that the field equations become second order around specific backgrounds. This might be however a problematic procedure to follow since, in essence, this construction amounts to requiring that those backgrounds correspond to surfaces in phase space, or in the space of solutions, where the principal part of the equations is singular, with all the associated problems that this brings about. For instance, standard perturbation theory will not be well-defined around such a singular surface or, at least, not in a standard form. A crucial feature to notice is that the principal part becomes singular not at a given point or surface in the spacetime (in which case regular solutions might still exist thanks to the mildness of the singular behaviour of the principal part in that situation), but on a surface in the space of solutions, which is a fundamentally more pathological situation. Thus, the very construction of the theory suggests that the backgrounds with the same linear spectrum as GR will be strongly coupled and, therefore, they cannot correspond to stable trajectories in phase space, i.e., no physical curves in phase space will smoothly evolve towards those solutions. Notice that this pathology is independent from the existing Ostrogradski instability in arbitrary higher order curvature theories and it would exist even if the full theory did not contain any ghosts. What this discussion suggests is that the strongly coupled modes will in turn be associated to the ghostly dof's present in the theory. 

It can be argued that the evanescence of dof's around those backgrounds is caused by some modes becoming infinitely massive so that they decouple. The reasoning to appreciate the strong coupling problem in this view is that the modes become infinitely massive after canonical normalisation. However, this canonical normalisation will also affect all the interaction terms and by the same token the effective couplings in the interactions will become infinite, i.e., the perturbations are infinitely strongly coupled. This is just another view on the problematic nature of singular surfaces of the principal part of the equations. Of course, it may happen that the full structure of the interactions at all orders is such that this problem never appears and then the infinitely massive case is not really problematic, precisely because it was never in the full spectrum of the theory. This happens for instance in the paradigmatic cases of $f(R)$ or $f(\mathcal{G})$ theories, with $R$ and $\mathcal{G}$ the Ricci scalar and the Gauss-Bonnet invariant respectively, that feature an additional scalar mode, whereas the extra spin-2 field of a general higher order curvature theory is not present. In other words, if we consider for instance the theory with $R^2+a R_{\mu\nu} R^{\mu\nu}$ and we take the limit $a\rightarrow0$ we will see that the mass of the extra spin-2 field goes to infinity (in appropriate variables) without incurring into any pathologies. Let us notice however that the limit is not continuous in the number of degrees of freedom so some care should be taken in performing the limit. In a general theory then, a background solution giving infinitely massive modes can signal the presence of pathologies and a careful analysis of the interactions (or of the full non-linear spectrum) must be performed to unveil its healthiness. 

The goal of the present paper will be to discuss in some detail the presence of the aforementioned pathologies for ECG. For the sake of simplicity, we will focus on cosmological scenarios (see e.g. \cite{cisterna2020four, Arciniega:2018fxj, Arciniega:2018tnn}) where the symmetries allow a more straightforward analysis of the pathologies, although, as we will discuss, our findings will be a general property of the theory around other backgrounds \cite{Hennigar2017, Bueno2016, Hennigar2017b, Bueno2017, Ahmed2017, Bueno2017b, Feng2017, Hennigar2017c, Hennigar2018, Bueno2019, Bueno2018, Poshteh2019, Mir2019,Mir2019b,Mehdizadeh2019,Erices2019}. Recently, it has been shown the presence of instabilities in these theories for inflationary solutions in \cite{Pookkillath:2020iqq}. We will confirm these findings and provide complementary evidence for the generic pathological character of these solutions. Furthermore, we will also consider the case of the generalised quasi-topological gravity theories (GQTG) introduced in \cite{Bueno:2019ycr} that extend the defining property of ECG to higher orders in curvature, although the particular form of the Lagrangian depends on the dimension. For these theories we will show that arbitrary cosmological solutions are even more pathological than the ECG case in a sense that we will make precise below.

The paper is organised as follows: In Section \ref{sec: ECG} we introduce the ECG Lagrangian and briefly comment on the potential pathologies of the theory. Then in Section \ref{sec: analysis ECG}, we analyse the behaviour of its equations of motion for Bianchi I solutions defined around an isotropic background and show from different points of view the singular behaviour of the dynamical equations. These instabilities are generically present in GQTG, as it is shown in Section \ref{sec: analysis quasitop} at least for the next three higher order Lagrangians. Finally we present and discuss our main results in Section \ref{sec: discussion}.

\noindent \bf{Conventions}: We will use the mostly plus signature for the metric, and define the Riemann tensor as $R_{\mu\nu\rho}{}^\lambda \equiv - 2 \big( \partial_{[\mu}\Gamma^\lambda_{\nu]\rho} + \Gamma^\lambda_{[\mu|\sigma} \Gamma^\sigma_{|\nu]\rho} \big)$ and the Ricci tensor is $R_{\mu\nu}\equiv R_{\mu\lambda\nu}{}^\lambda$.

\section{Einsteinian Cubic Gravity}\label{sec: ECG}

It will be convenient to begin by providing a brief introduction to ECG. We will be mainly interested in studying the cosmological behaviour of the theory so we will consider the version introduced in \cite{Arciniega:2018fxj} that extends the original ECG construction \cite{Bueno:2016xff} to arbitrary cosmological scenarios (FLRW). The main criterion followed in this construction was to obtain second order cosmological equations for the scale factor. This theory is described by the action
\begin{equation} \label{eq: action}
    \mS = \int \dd^4x \sqrt{|g|} \left(- \Lambda + \frac{\mpl^2}{2} R + \frac{\pECG}{\mpl^2} \mathcal{R}_{(3)} \right)\,,
\end{equation}
where the first two terms simply reproduce the pure GR sector (including a cosmological constant) and $\mathcal{R}_{(3)}$ is given by the following precise cubic polynomial of the Riemann tensor:
\begin{align}
    \mathcal{R}_{(3)} & = - \frac{1}{8} \Big(
    12 R_\mu{}^\rho{}_\nu{}^\sigma R_\rho{}^\tau{}_\sigma{}^\eta R_\tau{}^\mu{}_\eta{}^\nu  
    +R_{\mu\nu}{}^{\rho\sigma} R_{\rho\sigma}{}^{\tau\eta} R_{\tau\eta}{}^{\mu\nu}
    +2 R R_{\mu\nu\rho\sigma}R^{\mu\nu\rho\sigma}\nonumber \\
    & \qquad \qquad
    -8 R^{\mu\nu}R_\mu{}^{\rho\sigma \tau}R_{\nu\rho\sigma\tau}
    +4 R^{\mu\nu}R^{\rho\sigma}R_{\mu\rho\nu\sigma}
    -4 R R_{\mu\nu}R^{\mu\nu}
    +8 R_\mu{}^\nu R_\nu{}^\rho R_\rho{}^\mu
    \Big)\,.
\end{align}
The relative coefficients are carefully selected to guarantee that the linear spectrum of the theory around maximally symmetric and cosmological backgrounds is the same as that of GR, i.e., only the usual two polarisations of the gravitational waves propagate.  

For a Friedman-Lema\^itre-Robertson-Walker universe described by the line element
\be
\dd s^2=-\dd t^2+a^2(t) \dd \vec{x}^{\,2}\,, \label{eq:FLRW}
\ee
the gravitational Friedmann equation reads
\begin{equation}\label{eq: Isotropic condition}
3\mpl^2H^2(t)-6\frac{\pECG}{\mpl^2} H^6(t)=\rho+\Lambda\,,
\end{equation}
where $\rho$ is the energy-density of the matter sector. In the absence of any matter $\rho=0$, we can see that we have (at most) three branches of expanding de Sitter solutions. Among these de Sitter branches, the stability of the gravitational waves will impose some stability conditions. Indeed, if we consider metric perturbations $g_{ij}=a^2(\delta_{ij}+h_{ij})$ with $h_{ij}$ transverse and traceless, the corresponding quadratic action for the tensor perturbations around the de Sitter solutions is given by (see e.g. \cite{Pookkillath:2020iqq})
\be \label{Eq:actiontensormodes}
\mS^{(2)}= \frac{\mpl^4-6 H^4_0\pECG}{8 \mpl^2} \sum_{\lambda} \int \dd t \, \dd^3x \, a^3 \left[ \dot{h}^2_\lambda- \frac{1}{a^2} (\partial_i h_\lambda)^2\right],
\ee
where the sum extends to the two polarisations of the gravitational waves and $H_0$ is the considered de Sitter branch. From this expression it follows that we need to have $\mpl^4 - 6 H^4_0\pECG>0$ in order to avoid ghostly gravitational waves.

This quadratic action for the tensor modes explicitly shows the property of these theories that only the usual polarisations of the GWs propagate on a cosmological background at linear order. However, being a higher order curvature theory outside the Lovelock class, the full theory is expected to contain up to five degrees of freedom, which means that three of them will only enter at higher order in perturbation theory. As we will discuss in the following, precisely the trivialisation of these additional modes is at the heart of the pathological character of these solutions. Instead of working out the full non-linear terms, we will consider a simplified scenario that captures the non-linear origin of the pathologies in a simplified manner. 

\section{Approaching FLRW from Bianchi I in ECG} \label{sec: analysis ECG}
The singular nature of the FLRW solutions obtained in Einsteinian Cubic Gravity can be seen by breaking the isotropy of the spacetime while keeping homogeneity and study the evolution of the universe near the isotropic case. This analysis will be equivalent to considering a very long wavelength gravitational wave so it will serve as a proxy to illustrate the expected pathologies. In other words, a gravitational wave with a sufficiently long wavelength mimics the shear of a Bianchi I universe, so our study will give the non-linear evolution of gravitational waves, but restricted to the infrared sector.  Our analysis will partially overlap with the results presented in \cite{Pookkillath:2020iqq} and we will seize the opportunity to discuss some subtleties and give complementary arguments that will support the pathological character of these solutions.

\subsection{Bianchi I solutions. Theory-independent generalities}
Thus let us consider in principle an arbitrary gravitational action $\mS[g_{\mu\nu}]$, and the Bianchi I spacetime, described by the line element
\be
\dd s^2=-\mathcal{N}^{\,2}(t)\dd t^2+a^2(t) \dd x^2+b^2(t) \dd y^2+c^2(t) \dd z^2\,, \label{eq: BianchiI}
\ee
where $\mathcal{N}$ is the lapse function and $a$, $b$ and $c$ stand for the scale factors along the three coordinate axis. We will work in cosmic time so we will eventually make $\mathcal{N}(t)=1$, but we need to keep it general to properly compute all the gravitational equations. To make a more direct contact with the isotropic FLRW solutions, it is convenient to introduce the isotropic scale factor ${\bar a}\equiv (abc)^{1/3}$ with the corresponding expansion rate
\begin{equation}\label{eq: defH}
H(t)\equiv\frac{\dot{\bar a}}{\bar a}=\frac{1}{3}\left(\frac{\dot{a}}{a}+\frac{\dot{b}}{b}+\frac{\dot{c}}{c}\right)\,.
\end{equation}
In addition, we will encode the anisotropic part in two functions, $\sigma_1(t)$ and $\sigma_2(t)$, defined implicitly by \begin{equation} \label{eq: def sigmas}
\frac{\dot{a}}{a}= H + \eanis (2\sigma_1-\sigma_2)\,, \qquad 
\frac{\dot{b}}{b}= H - \eanis (\sigma_1-2\sigma_2)\,, \qquad 
\frac{\dot{c}}{c}= H - \eanis (\sigma_1+\sigma_2)\,.
\end{equation}
where $\eanis$ is certain (not necessarily small) fixed parameter that encapsulates the deviation with respect to the isotropic case given by $\eanis=0$. Notice that these definitions are consistent with \eqref{eq: defH}. 

Since the metrics of the type \eqref{eq: BianchiI} fulfill the requirements of the Palais' principle of symmetric criticality, one can use the minisuperspace approach and plug the Ansatz \eqref{eq: BianchiI} into the action before taking the variation. We introduce the convenient notation
\begin{align} 
  {\rm E}_{ab} \equiv \quad 0 & = \frac{\delta \bar{\mS}}{\delta a}a - \frac{\delta \bar{\mS}}{\delta b}b\,, & 
  {\rm E}_{cb} \equiv \quad 0 & = \frac{\delta \bar{\mS}}{\delta c}c - \frac{\delta \bar{\mS}}{\delta b}b\,,  & 
  {\rm E}_{ca} \equiv \quad 0 &=\frac{\delta\bar{\mS}}{\delta c}c-\frac{\delta\bar{\mS}}{\delta a}a\,,\nonumber \\
  {\rm E}_\mathcal{N} \equiv \quad 0 & =\frac{\delta \bar{\mS}}{\delta \mathcal{N}}\,, &
  {\rm E}_{abc} \equiv \quad 0 & =\frac{1}{3} \left(\frac{\delta\bar{\mS}}{\delta a}a+\frac{\delta\bar{\mS}}{\delta b}b+\frac{\delta\bar{\mS}}{\delta c}c\right)\,.
\end{align}
where $\bar{\mS}$ is the gravitational action $\mS$ evaluated in the Ansatz \eqref{eq: BianchiI}. However, not all of them are independent equations, because $\{{\rm E}_{abc}, {\rm E}_{ab}, {\rm E}_{cb},  {\rm E}_{ca} \}$ are linearly dependent and the Bianchi identity associated to diffeomorphisms,
\begin{equation}
  \frac{\dd}{\dd t}\left(\frac{\delta \bar{\mS}}{\delta \mathcal{N}}\right) + 
  \left(\frac{\dot{\mathcal{N}}}{\mathcal{N}}+ \frac{\dot{a}}{a}+\frac{\dot{b}}{b}+\frac{\dot{c}}{c} \right)\frac{\delta \bar{\mS}}{\delta \mathcal{N}} - \frac{1}{\mathcal{N}}
  \left(\dot{a}\frac{\delta \bar{\mS}}{\delta a} +\dot{b}\frac{\delta \bar{\mS}}{\delta b} + \dot{c}\frac{\delta \bar{\mS}}{\delta c}\right)=0\,,
\end{equation}
establishes an additional relation among them. The particular parameterisation of the two shears $\sigma_{1,2}$ introduced in \eqref{eq: def sigmas} has been chosen conveniently to work with $\{{\rm E}_\mathcal{N},  {\rm E}_{cb},  {\rm E}_{ca}\}$. However, in order to obtain a more direct generalisation of the results obtained in \cite{Pookkillath:2020iqq}, only in section \ref{subsec: perturbative sol} we will make the (equivalent) choice $\{{\rm E}_\mathcal{N},  {\rm E}_{ab},  {\rm E}_{ca}\}$.

\subsection{Perturbative solution around de Sitter spacetime} \label{subsec: perturbative sol}
In  \cite{Pookkillath:2020iqq} it was argued that de Sitter is a stable perturbative solution of \eqref{eq: action} in the case of an axisymmetric Bianchi I. We will reproduce here the same analysis for the general Bianchi I case in order to clarify some subtle shortcomings of the solutions generated perturbatively. 

Trivially from \eqref{eq: Isotropic condition}, the de Sitter spacetime given by
\begin{equation}
  \mathcal{N}(t)=1\quad\text{and}\quad a(t)=b(t)=c(t)=\eN^{H_0t}
\end{equation}
is a solution of $\{{\rm E}_\mathcal{N},  {\rm E}_{ab},  {\rm E}_{ca}\}$ if and only if the cosmological constant and the Hubble
parameter fulfill
\begin{equation}
  \Lambda=3\frac{H_0^2}{\mpl^2} (\mpl^4-2H_0^4\pECG)\,.\label{eq: L H0 relation}
\end{equation}
Now consider a perturbative expansion around this isotropic configuration parameterised as follows:
\begin{equation}
  a(t) =a^{(0)}(t)+ \sum_{k=1}^\infty \epsilon^k a^{(k)}(t)\,, \qquad
  b(t) =a^{(0)}(t)+ \sum_{k=1}^\infty \epsilon^k b^{(k)}(t)\,, \qquad
  c(t) =a^{(0)}(t)+ \sum_{k=1}^\infty \epsilon^k c^{(k)}(t)\,,
  \label{Eq:perturbativeAnsatz}
\end{equation}
where $a^{(0)}(t)\equiv\eN^{H_0 t}$ and $\epsilon$ is the small perturbation parameter (not to be confused with $\epsilon_{\sigma}$ introduced above). If we substitute the perturbative expansion \eqref{Eq:perturbativeAnsatz} into the dynamical equations $\{{\rm E}_\mathcal{N},  {\rm E}_{ab},  {\rm E}_{ca}\}$, and since the background is a solution, we expect the first non-trivial contribution to appear at first order. Indeed, the resulting equations are
\begin{align}
  0 & = \dot{a}^{(1)} + \dot{b}^{(1)}+ \dot{c}^{(1)}- H_0 \big(a^{(1)}+b^{(1)}+c^{(1)}\big)  \,,\nonumber \\
  0 & =  \ddot{b}^{(1)} - \ddot{a}^{(1)}+ H_0 \big(\dot{b}^{(1)} - \dot{a}^{(1)}\big) - 2H_0^2 \big(b^{(1)} -a^{(1)}\big) \,,\nonumber \\
  0 & =\ddot{c}^{(1)} - \ddot{a}^{(1)}+ H_0 \big(\dot{c}^{(1)} - \dot{a}^{(1)}\big) - 2H_0^2 \big(c^{(1)} -a^{(1)}\big)\,,
  \label{eq:firstorder}
\end{align}
provided $\mpl^4 \neq 6 H_0^4\pECG$. As shown in \eqref{Eq:actiontensormodes}, this is in turn a necessary condition to avoid a pathological behaviour of the tensor modes, which requires $\mpl^4 - 6 H_0^4\pECG>0$ (see also \cite{Pookkillath:2020iqq}). The general solution for this first order contribution is given by the expressions
\begin{align}
  a^{(1)} & =C_1\eN^{-2H_0t}+C_3\eN^{H_0t}\,,\nonumber \\
  b^{(1)} & =C_2\eN^{-2H_0t}+C_4\eN^{H_0t}\,,\nonumber \\
  c^{(1)} & =-(C_1+C_2)\eN^{-2H_0t}+C_5 \eN^{H_0t}\,,
\end{align}
for some integration constants $C_i$ ($i=1,...,5$). 
We can proceed analogously to obtain the solution at second order that is found to be
\begin{align}
a^{(2)} & = 
  \frac{C_1^2-C_1C_2-C_2^2}{4}\frac{\mpl^4-258H_0^4\pECG}{\mpl^4-6H_0^4\pECG}\eN^{-5H_0t}+
  \frac{1}{3}\big[\bar{C}+D_1\big]\eN^{-2H_0t}+
  D_3\eN^{H_0t}\,,\nonumber \\
b^{(2)} & =
  \frac{C_2^2-C_1C_2-C_1^2}{4}\frac{\mpl^4-258H_0^4\pECG}{\mpl^4-6H_0^4\pECG}\eN^{-5H_0t}+
  \frac{1}{3}\big[\bar{C}+D_2\big]\eN^{-2H_0t}+
  D_4\eN^{H_0t}\,,\nonumber \\
c^{(2)} & =
  \frac{C_1^2+3C_1C_2+C_2^2}{4}\frac{\mpl^4-258H_0^4\pECG}{\mpl^4-6H_0^4\pECG}\eN^{-5H_0t}+
  \frac{1}{3}\big[\bar{C}-D_1-D_2\big]\eN^{-2H_0t}+
  D_5\eN^{H_0t}\,.
\end{align}
where $D_{i}$ ($i=1,...,5$) are new integration constants and $\bar C\equiv C_1C_3+C_2C_4-(C_1+C_2)C_5 $. 

The resulting perturbative expansion reproduces the exactly isotropic de Sitter solution at all orders, since the anisotropic contributions decay exponentially so that the isotropic de Sitter solution is reached exponentially fast. Notice that the perturbative contributions proportional to $\eN^{H_0 t}$ can be absorbed into the background solution. This behaviour was argued in \cite{Pookkillath:2020iqq} to guarantee the existence of FLRW solutions. However, some care must be taken to correctly interpret this perturbative solution since the zeroth order corresponds to a singular surface in phase space where dynamical degrees of freedom disappear, as we will see in the next section. Thus, performing a standard perturbative expansion around this surface can be problematic and the conclusions drawn from it can be flawed. In the present case, there is a hint about the ill-defined nature of the perturbative expansion in the fact that each order is obtained by solving second order equations, while the full equations are known to be fourth order. In (\ref{eq:firstorder}) we can explicitly see the absence of the higher than second order derivatives in the equations at first order, but it is easy to understand that this will be the case at all orders. The reason is that the coefficients of the terms with third and fourth  derivatives of the scale factors at $n$-th order must be evaluated on the purely isotropic zeroth order solution. Since the isotropic case, by definition of the theory, gives second order equations, these coefficients must vanish. In other words, the terms with higher derivatives at $n$-th order will always start at $(n+1)$-th order. In the general anisotropic case studied in next section, we will demonstrate this property at the full non-linear level. The importance of this observation is that it implies that we are necessarily missing perturbative modes along specific directions whose stability is not under control and, certainly, they are not captured by the generated perturbative solution. Thus, we cannot conclude that the de Sitter solution is a good background solution from the above perturbative analysis. In Appendix \ref{app: toyexample} we illustrate these issues with a simple one-dimensional mechanical toy example where the explained problematic nature of the perturbative solution around a singular surface in phase space is clearly visible.

\subsection{Shear equations for Einsteinian Cubic Gravity}

In the theory \eqref{eq: action}, one finds that the highest order derivatives of the anisotropy functions $\sigma_1$ and $\sigma_2$ appear in the evolution equations (shear equations), ${\rm E}_{cb}$ and ${\rm E}_{ca}$. Obviously, these equations trivialise in the isotropic case because they describe the evolution of the shear $\sigma_1$ and $\sigma_2$, which means that there is an overall factor $\eanis$. In the absence of any anisotropic stress, as we are considering, and away from the isotropic case, the shear evolution equations can be taken to be $\{{\rm E}_{cb}/\eanis, {\rm E}_{ca}/\eanis\}$, which can be written in the following schematic form:
\renewcommand\arraystretch{1.3}
\begin{equation}\label{eq: principal part}
  \frac{\pECG}{\mpl^2} \left[ \eanis
    {\bf M_1}
    \begin{pmatrix}\dddot{\sigma}_2 \\\dddot{\sigma}_1 \end{pmatrix}
    + \eanis 
    {\bf M_2} \begin{pmatrix}\ddot{\sigma}_2 \\\ddot{\sigma}_1 \end{pmatrix}+  {\bf V} \right] 
    +3 \mpl^2 \begin{pmatrix}3H\sigma_2+\dot{\sigma}_2 \\ 3H\sigma_1+\dot{\sigma}_1 \end{pmatrix} =0 \,.
\end{equation}
\renewcommand\arraystretch{1}
where the matrices ${\bf M}_{1,2}$ and the column vector ${\bf V}$ start at zeroth order in $\eanis$. The components of ${\bf M}_1$ depend polynomially on $\sigma_1$, $\sigma_2$ and $H$, whereas those of ${\bf M}_2$ and ${\bf V}$ also depend on $\dot{\sigma}_1$, $\dot{\sigma}_2$ and the derivatives of $H$. This equation clearly shows how the higher order terms containing second and third derivatives of the shear trivialise in the isotropic limit $\eanis\rightarrow0$ so that the order of the corresponding differential equations is reduced. Notice that this property does not imply that the shear evolution is not modified by the ECG term in the action, since the usual GR evolution (described by the last term in the LHS of \eqref{eq: principal part}) receives corrections from ${\bf V}$. From this equation we can also understand the missing modes in the perturbative solution obtained in the previous section. If we consider a perturbative expansion $\sigma_{1,2}=\sum_{i=1}^\infty \sigma
^{(i)}_{1,2}$, it is apparent that the higher order derivative terms of the anisotropic modes $\sigma_{1,2}$ at $n$-th order will only contribute to $(n+1)$-th order because the coefficients of $\ddot{\sigma}_{1,2}$ and $\dddot{\sigma}_{1,2}$ start at first order, i.e., they trivialise in the isotropic case. 

On the other hand, having obtained the non-linear equation for the anisotropic homogeneous modes, we corroborate that the isotropic solution lies on a singular surface of phase space. This means that solutions near the singular isotropic surface can never end in the isotropic solution. At best, a given trajectory could approach the isotropic solution, but its intrinsically singular nature prevents the possibility of making any reliable claim. In particular, this is the reason why the perturbative expansion of Sec. \ref{subsec: perturbative sol} fails to capture the full perturbative spectrum around the isotropic solution. 

\subsection{Complete dynamical analysis}

In order to go deep into the pathological character of the isotropic solutions we can consider the full system of equations conformed by the shear equations \eqref{eq: principal part} together with the lapse equation ${\rm E}_\mathcal{N}$ that constitute the complete dynamical description of the theory in terms of $\sigma_1$, $\sigma_2$ and $H$ (we take $\eanis=1$ from now on). These equations can be recast as a system with up to third order time derivatives of $\sigma_1$ and $\sigma_2$ and up to second order time derivatives of $H$. Although the shear equations contain third order derivatives of $H$, they can be eliminated by taking successive time derivatives of ${\rm E}_\mathcal{N}$. This procedure results in additional corrections to the coefficients of the third order derivatives of the shear. We are interested in obtaining the matrix of the principal part of the equations once they are written in the discussed normal form, i.e., with only up to second derivatives of $H$. It is then convenient to factor the isotropic expansion out by introducing the variables
$X(t)\equiv \sigma_1/H(t)$ and $Y(t)\equiv\sigma_2/ H(t)$ and work with the number of e-folds
\begin{equation}  \dd N=H(t)\dd t \label{eq: N efolds}\end{equation}
as time variable (we will use a prime to represent the derivative with respect to $N$). After these manipulations, the full system of equations can be written as
\be \label{eq: systemeqs}
\mathcal{H}^i{}_j (\xi^j)' + F^i(X,Y,X',Y',X'',Y'',H,H')=0\,,
\ee
where $\vec{\xi}\equiv(H',X'',Y'')$, $\vec{F}$ is a vector that depends on the displayed dynamical variables and $\mathcal{H}^i{}_j$ is the desired matrix of coefficients for the principal part whose determinant is given by
\begin{equation}\label{eq: detHess}
     \det \mathcal{H}^i{}_j= - \frac{1458 \pECG^3 H^{15}}{\mpl^{12}} \big(X^2-2Y^2+2Y-X+2XY\big)\big(Y^2-2X^2+2X-Y+2XY\big)\big(Y^2+X^2-4XY-X-Y\big)\,.
\end{equation}

Besides the singular curves given by $\det \mathcal{H}^i{}_j=0$, there is an additional separatrix associated to a null eigenvalue along the $H''$ direction given by
\begin{equation}\label{eq: extrasep}
    2\big(X^3+Y^3\big)- 3\big(XY^2+X^2Y\big) -2\big(X^2+Y^2\big)+2XY=0\,.
\end{equation}
This separatrix does not appear from the vanishing of the determinant because the other two eigenvalues diverge on this curve in such a way that the determinant remains finite. Thus, we will also consider this separatrix in our analysis.
Notice that the expressions from the separatrices in \eqref{eq: detHess} and \eqref{eq: extrasep} are invariant under  $X\leftrightarrow Y$ (i.e., $\sigma_1\leftrightarrow\sigma_2$) as it should be, and only depend algebraically on $X$, $Y$. This means that these singular curves actually correspond to singular hypersurfaces in phase space that are orthogonal to the plane $(X,Y)$ or, equivalently, parallel to the other directions\footnote{Let us be more explicit on this and notice that the considered phase space is conformed by the coordinates $(H,X,Y,H',X',Y',X'',Y'')$. The singular surfaces then exhibit a  symmetry under translations in the subspace $(H,H',X',Y',X'',Y'')$.}.

\begin{figure} 
\includegraphics[width=0.4\textwidth]{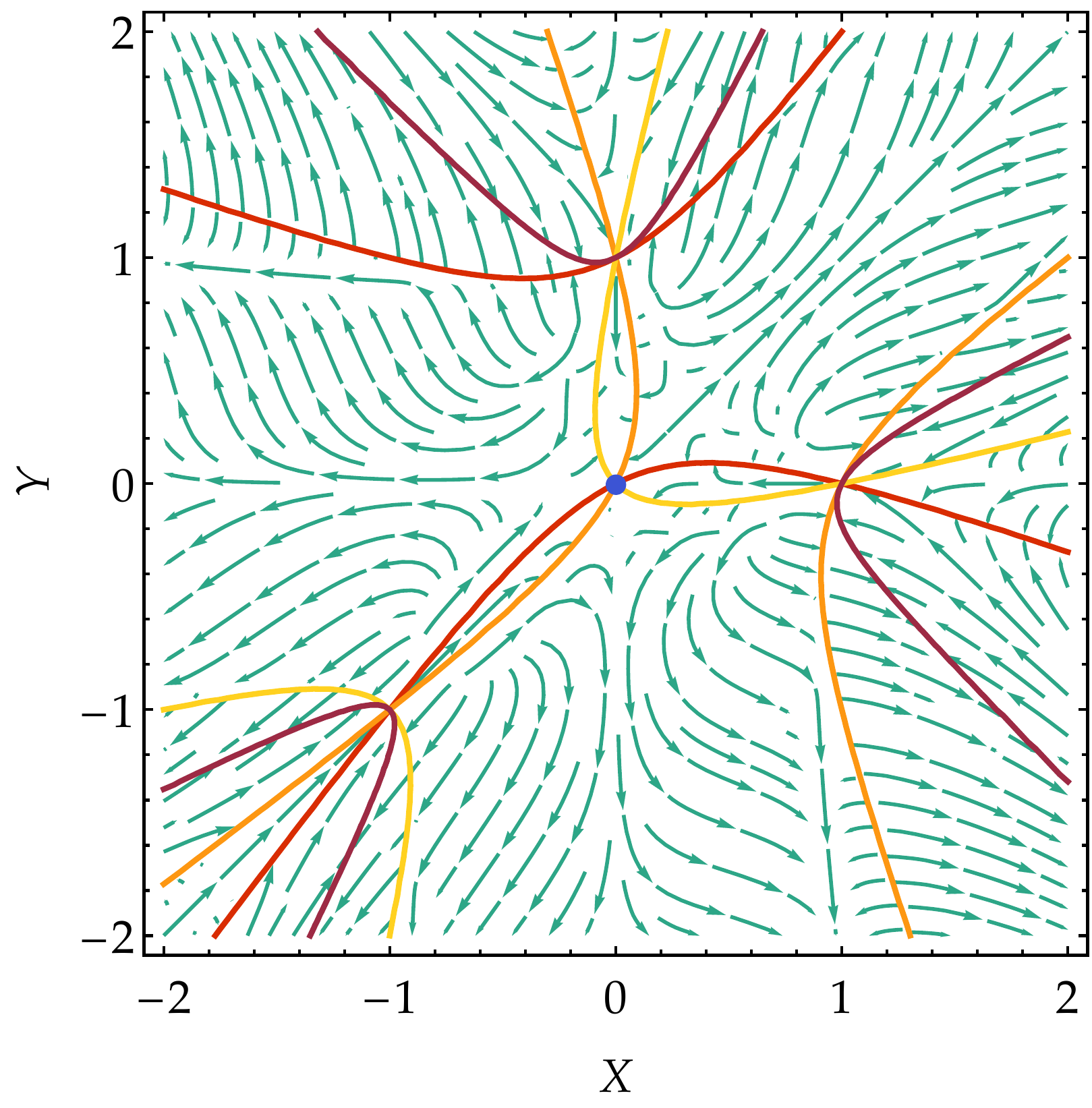} \qquad
\includegraphics[width=0.42\textwidth]{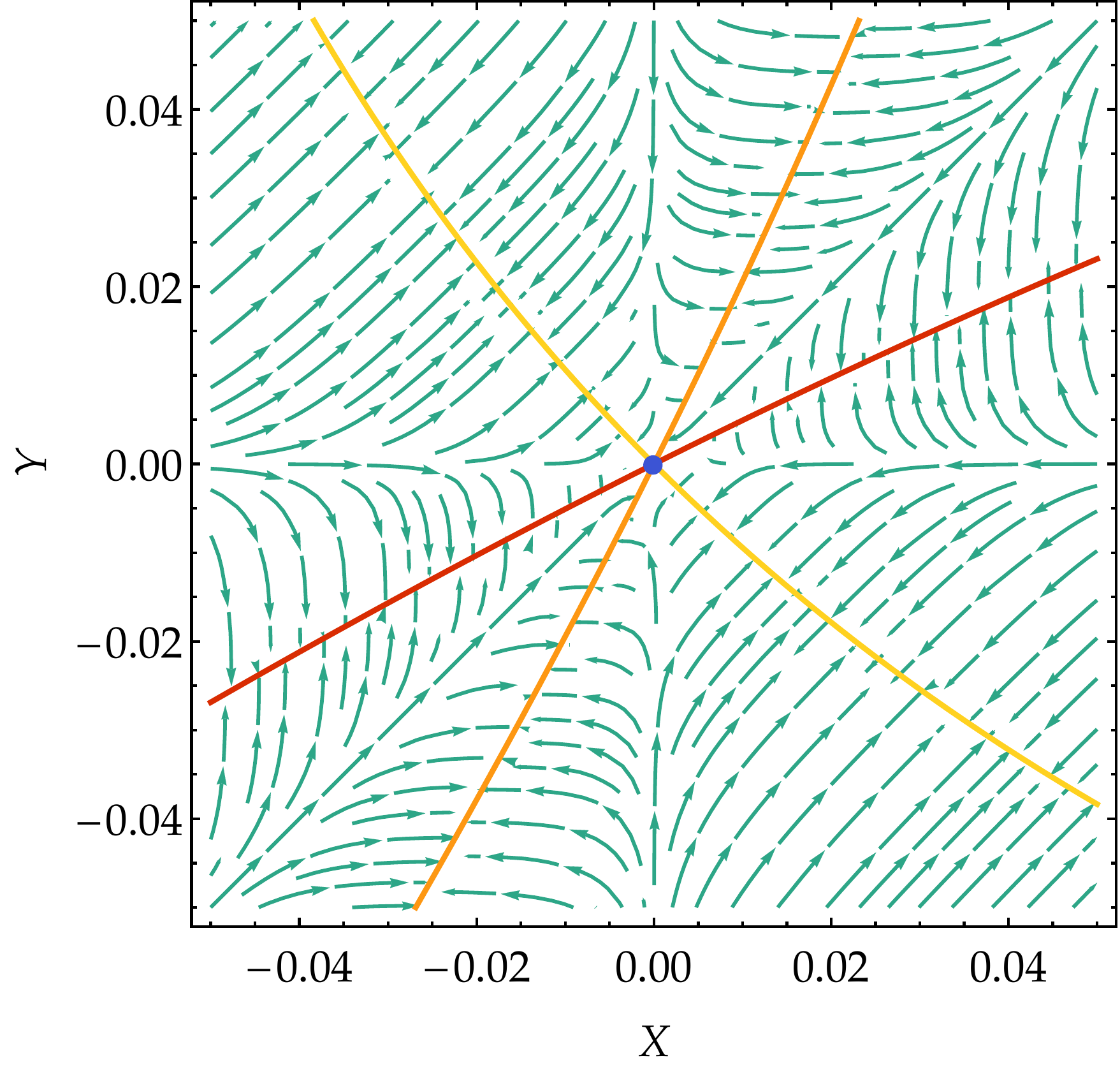}
\quad
\includegraphics[width=0.4\textwidth]{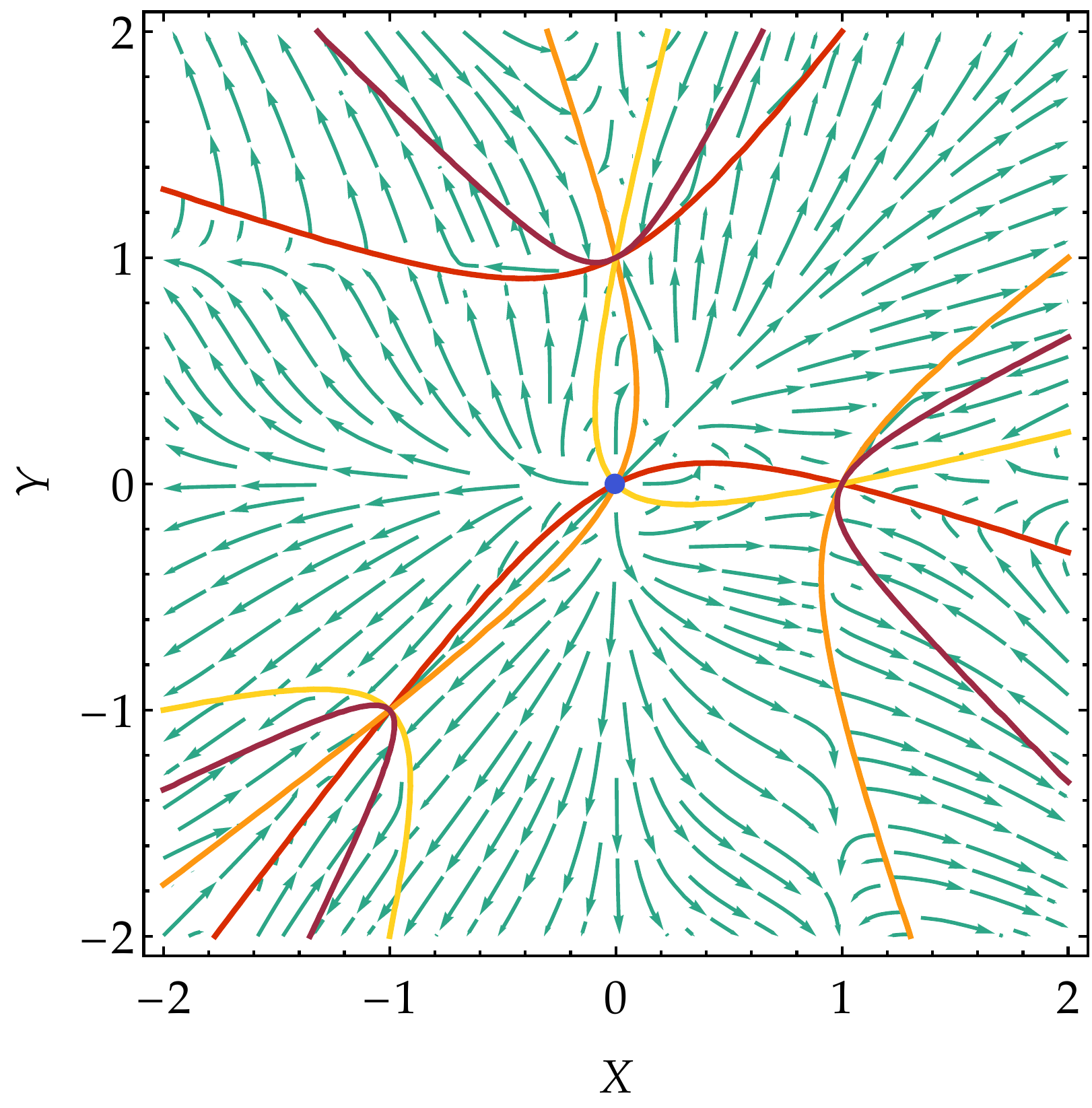}\qquad
\includegraphics[width=0.42\textwidth]{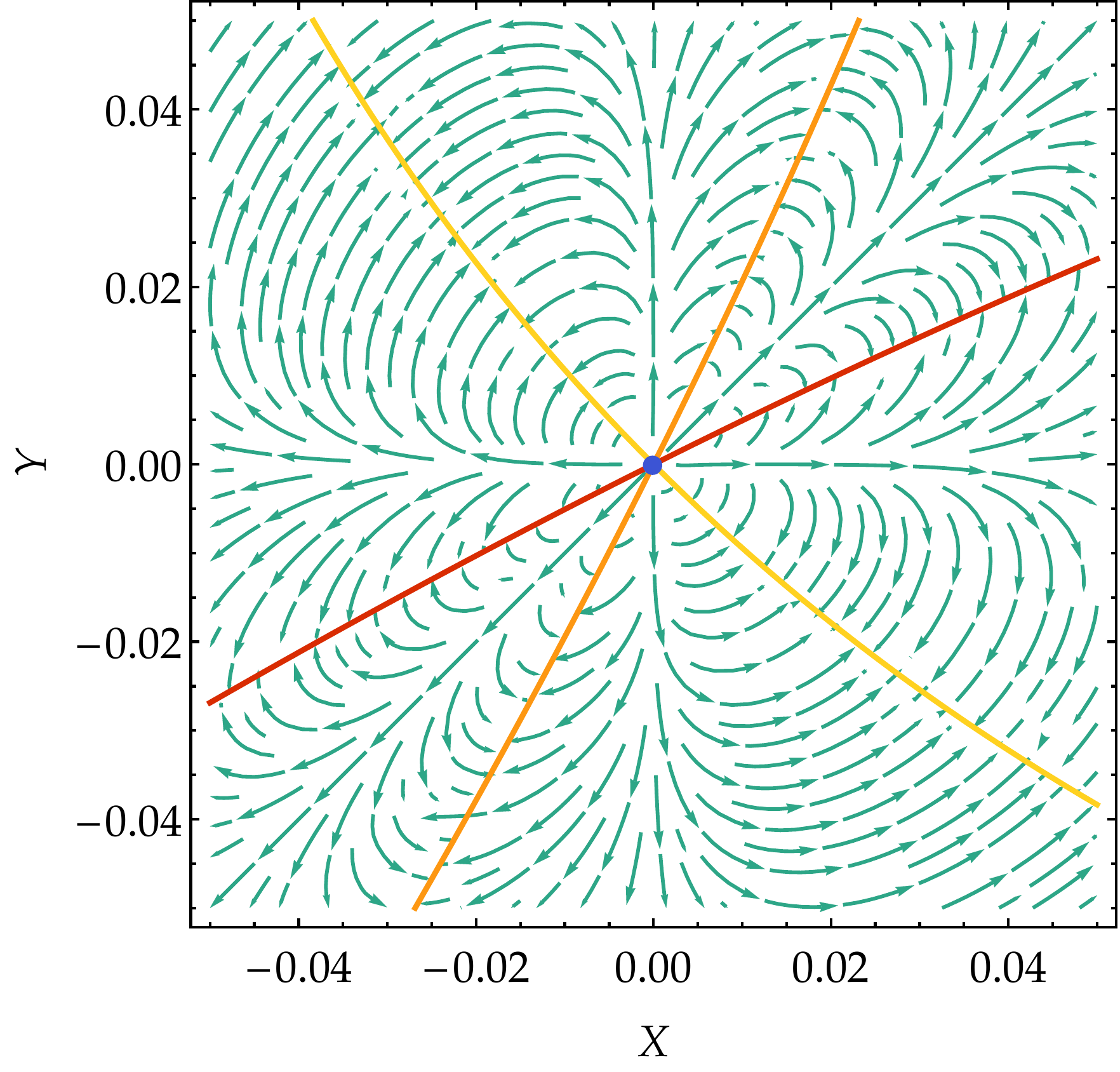}

\caption{\label{fig:phasemapECG}
These plots show how the field $(\xi^2,\,\xi^3)'=(X''',Y''')$ behaves with respect to $X$ and $Y$ under the conditions $H'=X'=Y'=X''=Y''=0$ and $H=H_\text{dS}$. The curves in red, orange and yellow are, respectively, the three branches of singular separatrices as they appear in \eqref{eq: detHess}, while the dark red one is the additional separatrix that does not appear in the Hessian determinant. The isotropic solution corresponds to the blue point at the origin. We can also see other three distinctive singular points that correspond to the physical solutions where only one of the directions expands (see Table \ref{tab: intersections}). The first two plots correspond to the value $\pECG \mpl^4/H^4_\text{dS} = 0.1$ and the last ones to $\pECG \mpl^4/H^4_\text{dS} = 0.001$. 
}
\end{figure}

In Fig. \ref{fig:phasemapECG}, the curves displayed in yellow, orange and red are the critical curves where the determinant \eqref{eq: detHess} vanishes, while the dark red ones represent the additional separatrix given by  \eqref{eq: extrasep}.
Notice that the isotropic point ($X=Y=0$) is crossed by the first three, whereas it is just an isolated solution of \eqref{eq: extrasep}. In total, the separatrices have only four intersection points as can be seen in Fig. \ref{fig:phasemapECG}. These special points are collected in Table \ref{tab: intersections} and they correspond, in addition to the isotropic solution, to the three FLRW universes with flat spatial slices in which only one of the spacelike directions is dynamical.

The analysis of the full dynamical system and the associated properties of the corresponding phase map is quite contrived. However, we do not need to perform a detailed analysis of such properties for our purposes, but we are only interested in explicitly showing the pathological character of the isotropic solutions. This should already be clear from the fact that such a solution in fact belongs to the discussed separatrices. We can show it in an even more explicit manner by considering the following restricted situation. First, since the separatrices only depend on $X$ and $Y$ we will focus on this plane of the phase space. Then we will consider the flow of the trajectories with $H'=X'=Y'=X''=Y''=0$ and $H=H_\text{dS}$ that correspond to trajectories that are anisotropically displaced from the de Sitter solution and left {\it at rest}. We can then plot the flow of the vector field $(\xi^2,\,\xi^3)'=(X''',Y''')$ in the $(X,Y)$ plane as we show in Fig. \ref{fig:phasemapECG}. To be clear, we are considering the phase space flow on the hypersurface $H'=X'=Y'=X''=Y''=0$ and $H=H_\text{dS}$, then we project onto the plane $(\xi^2,\,\xi^3)=(X'',Y'')$ and, finally, we plot how this projected flow varies with the coordinates $(X,Y)$ (that can be interpreted as external parameters for the resulting vector field). A cautionary word might be in order here. The plots in Fig. \ref{fig:phasemapECG} involve an identification of the $(X,Y)$ axis with the directions of $(\xi^2,\,\xi^3)$. Thus, although these diagrams provide a limited information on the physical trajectories, they can be used to clearly see the obtained separatrices as well as the crucial consequence that no physical solutions can smoothly approach them. In particular, we can see how the origin (that describes the isotropic solution) is an unstable point. 

After analytically showing the pathological nature of the isotropic solutions, in the next section we will provide some numerical examples to clearly illustrate the given general arguments.

\begin{table}
  \renewcommand\arraystretch{1.6}
  \begin{tabular}{|c|c|c||l|}
    \hline 
    $(X,\,Y)$ & $(\sigma_{1},\,\sigma_{2})$ & $\left(\frac{\dot{a}}{a},\,\frac{\dot{b}}{b},\,\frac{\dot{c}}{c}\right)$ & Description \tabularnewline
    \hline \hline 
    $(0,\,0)$ & $(0,\,0)$ & $\left(H,\,H,\,H\right)$ & Isotropic point \tabularnewline
    \hline 
    $(1,\,0)$ & $(H,\,0)$ & $\left(3H,\,0,\,0\right)$ & $b, c$ constant functions \tabularnewline
    \hline 
    $(0,\,1)$ & $(0,\,H)$ & $\left(0,\,3H,\,0\right)$& $a, c$ constant functions \tabularnewline
    \hline 
    $(-1,\,-1)$ & $(-H,\,-H)$ & $\left(0,\,0,\,3H\right)$& $a, b$ constant functions \tabularnewline
    \hline 
  \end{tabular}
  \renewcommand\arraystretch{1}
  
  \caption{\label{tab: intersections}
    In this table we summarise the four special points in the plane $XY$ where the different singular branches intersect. These special solutions correspond to universes with isotropic evolution and with evolution along one of the directions while the transverse one remain static.}
  
\end{table}

\subsection{Numerical analysis}
\begin{figure} 
\includegraphics[width=0.4\textwidth]{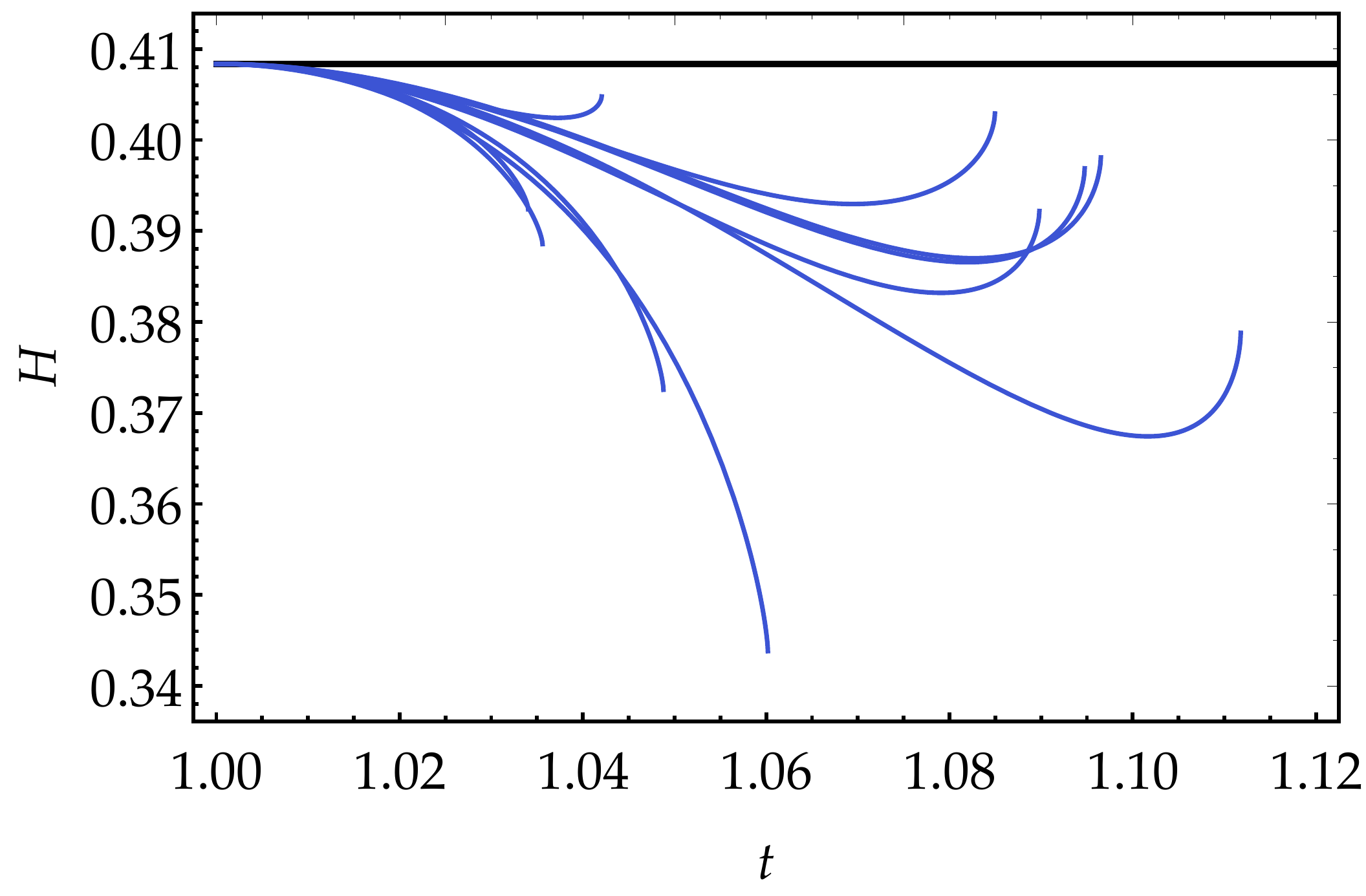} \qquad
\includegraphics[width=0.42\textwidth]{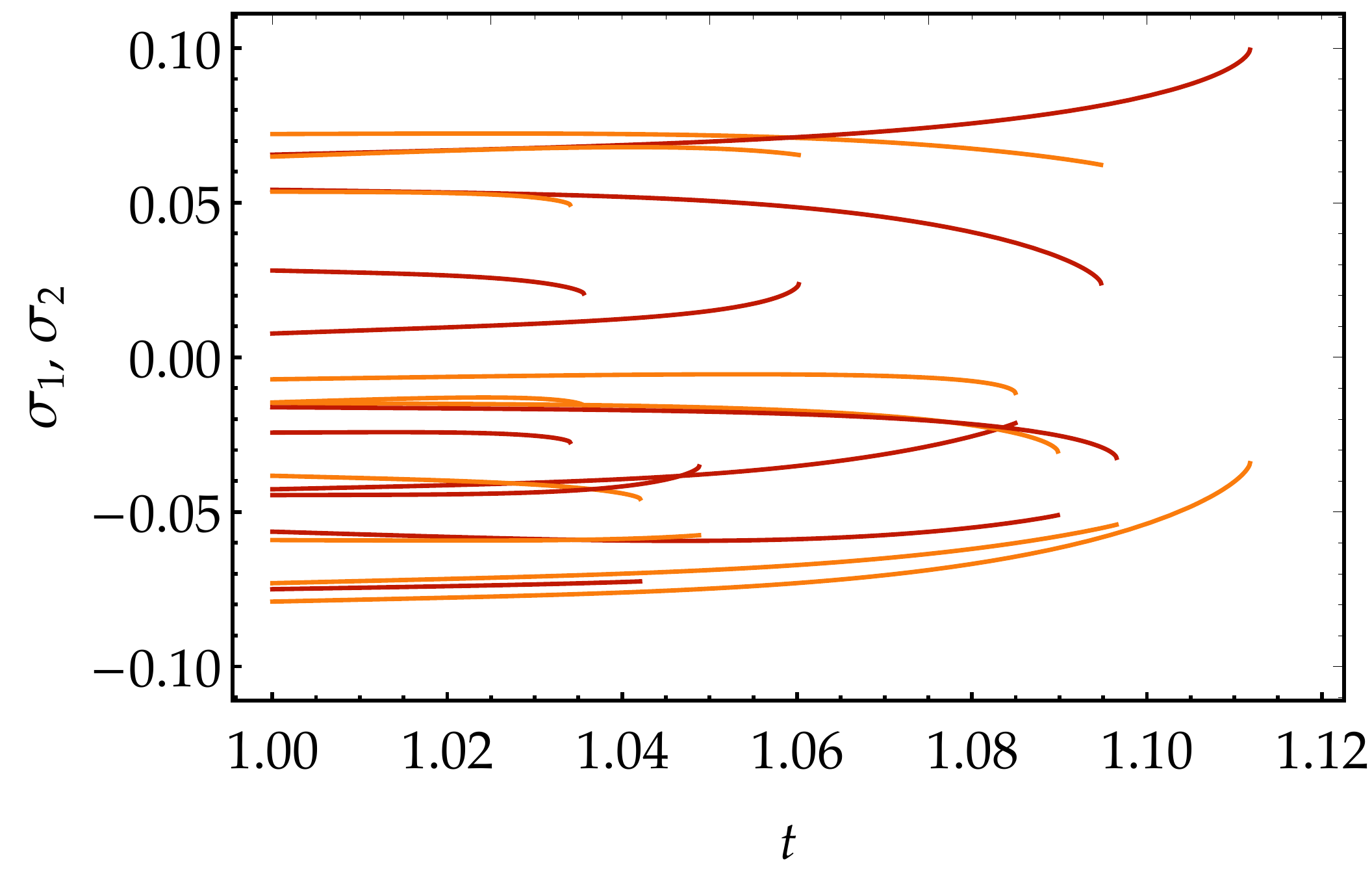}

\caption{
In the first plot we represent the isotropic solution for $H(t)$ (black line) and the numerical ones (blue lines) starting with the same initial conditions for $H$, but for randomly generated initial conditions for $\sigma_1(t)$ and $\sigma_2(t)$. The second plot shows the result of the integration for $\sigma_1(t)$ and $\sigma_2(t)$. For this numerical result we have chosen $\Lambda=0.5$, $\mpl=1$, $\pECG=0.01$ and the initial value $H(t=1)=0.408362$ (the only solution of \eqref{eq: Isotropic condition} with $2H^2<\Lambda$).}
\label{Fig:plotdeSitter}
\end{figure}

We will now examine numerical solutions for the full set of equations in the Bianchi I spacetime. We will use the independent set of differential equations corresponding to $\{{\rm E}_\mathcal{N}, {\rm E}_{cb}, {\rm E}_{ca} \}$. These give second order differential equations for the isotropic Hubble expansion rate $H$ and third order differential equations for $\sigma_{1,2}$. In the equations for the shear we replace $\ddot{H}$ and $\dddot{H}$ by the expressions obtained by taking successive time derivatives of ${\rm E}_\mathcal{N}$. For the initial conditions, we choose the de Sitter solution as our baseline and give perturbed initial conditions around this solution with a random but small amplitude to the initial shears and their derivatives. This procedure allows us to scan the phase space around the de Sitter solution. We show the obtained numerical solutions in Fig. \ref{Fig:plotdeSitter}, which confirms our discussion above. In the left panel we show the evolution for $H$ together with the exact de Sitter solution. We see that in all cases, the perturbed solutions quickly deviate from the isotropic one and that the solutions that eventually turn and approach the isotropic solution encounter a singular point beyond which the evolution ceases. Let us stress that this behaviour clearly reflects the fact that the solution reaches a singular point. For illustrative purposes we give an example with only ten solutions, but we have checked that this is the general tendency when more trajectories are explored.

\subsection{Numerical analysis for a radiation dominated era}

In this section we will perform a similar numerical integration but in the presence of a matter sector $\mS_{\rm matt}$ describing radiation, i.e., one whose energy-momentum tensor has the form
\begin{equation}
T^{\mu\nu}\equiv\frac{2}{\sqrt{|g|}}\frac{\delta \mS_{\rm matt}}{\delta g_{\mu\nu}}=(\rho_{\rm r}+P_{\rm r})u^\mu u^\nu+P_{\rm r}g^{\mu\nu}\,,
\end{equation}
where $P_{\rm r}=\frac{1}{3}\rho_{\rm r}$ and $u^\mu$ is the fluid 4-velocity. Therefore, among $\{{\rm E}_\mathcal{N},  {\rm E}_{cb},  {\rm E}_{ca}\}$, only the equation of the lapse is modified, according to $ {\rm E}_{\mathcal{N}} \to {\rm E}_{\mathcal{N}}-\rho_{\rm r}(t)$. 
In the presence of radiation it is convenient to work in terms of the number of e-folds $N$, defined in \eqref{eq: N efolds}. Then, the Bianchi identity associated to diffeomorphisms for the matter action,
\begin{equation}\label{eq: Bianchi r}
  0=\nabla_{\mu}T^{\mu\nu}\qquad\Rightarrow\qquad\frac{\dot{\rho}_{\rm r}(t)}{\rho_{\rm r}(t)}=-4H(t)\,,
\end{equation}
can be immediately integrated:
\begin{equation} \rho_{\rm r}(N)=\rho_0 \eN^{-4N}\,.\end{equation}

\begin{figure}

\includegraphics[width=0.4\textwidth]{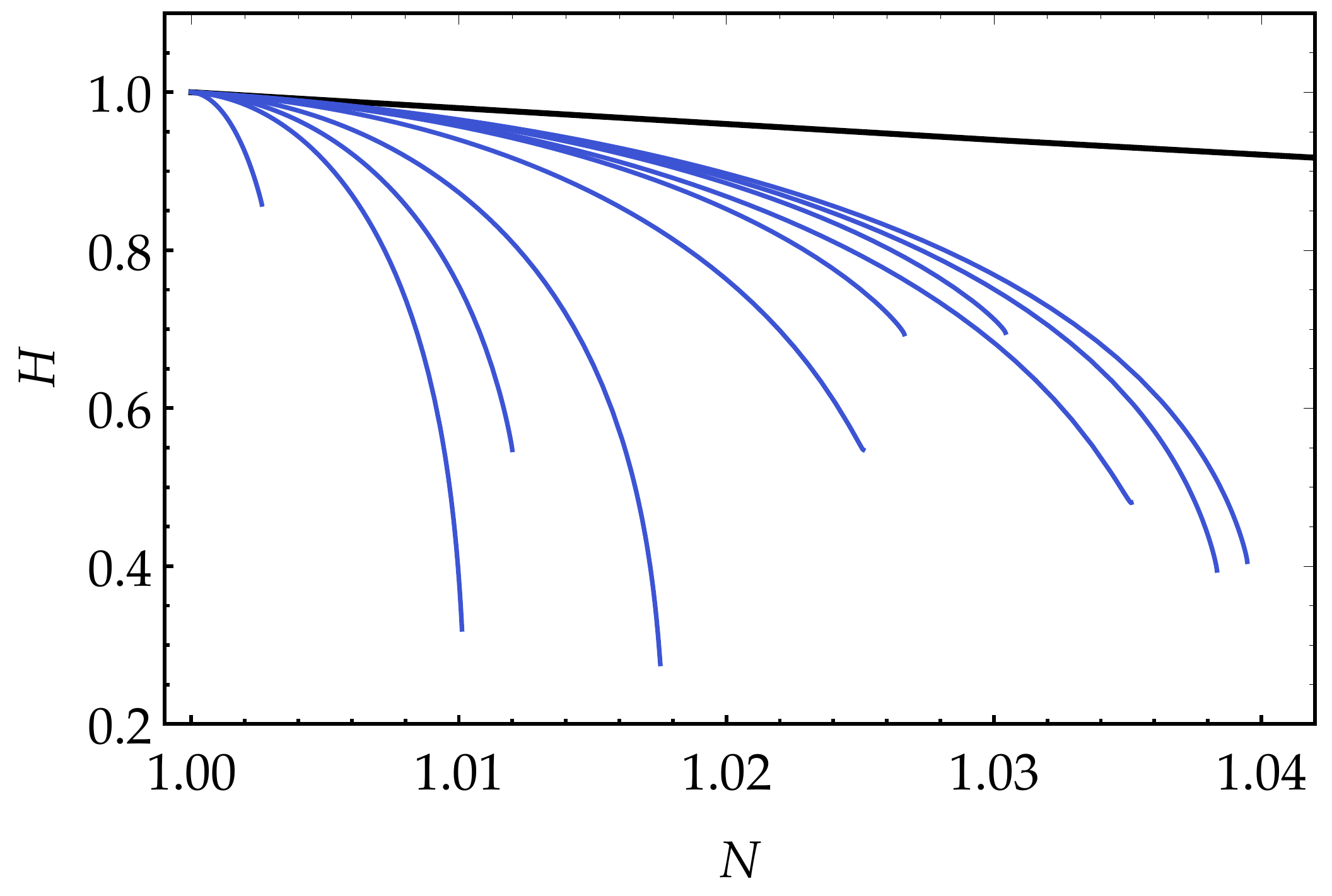} \qquad
\includegraphics[width=0.42\textwidth]{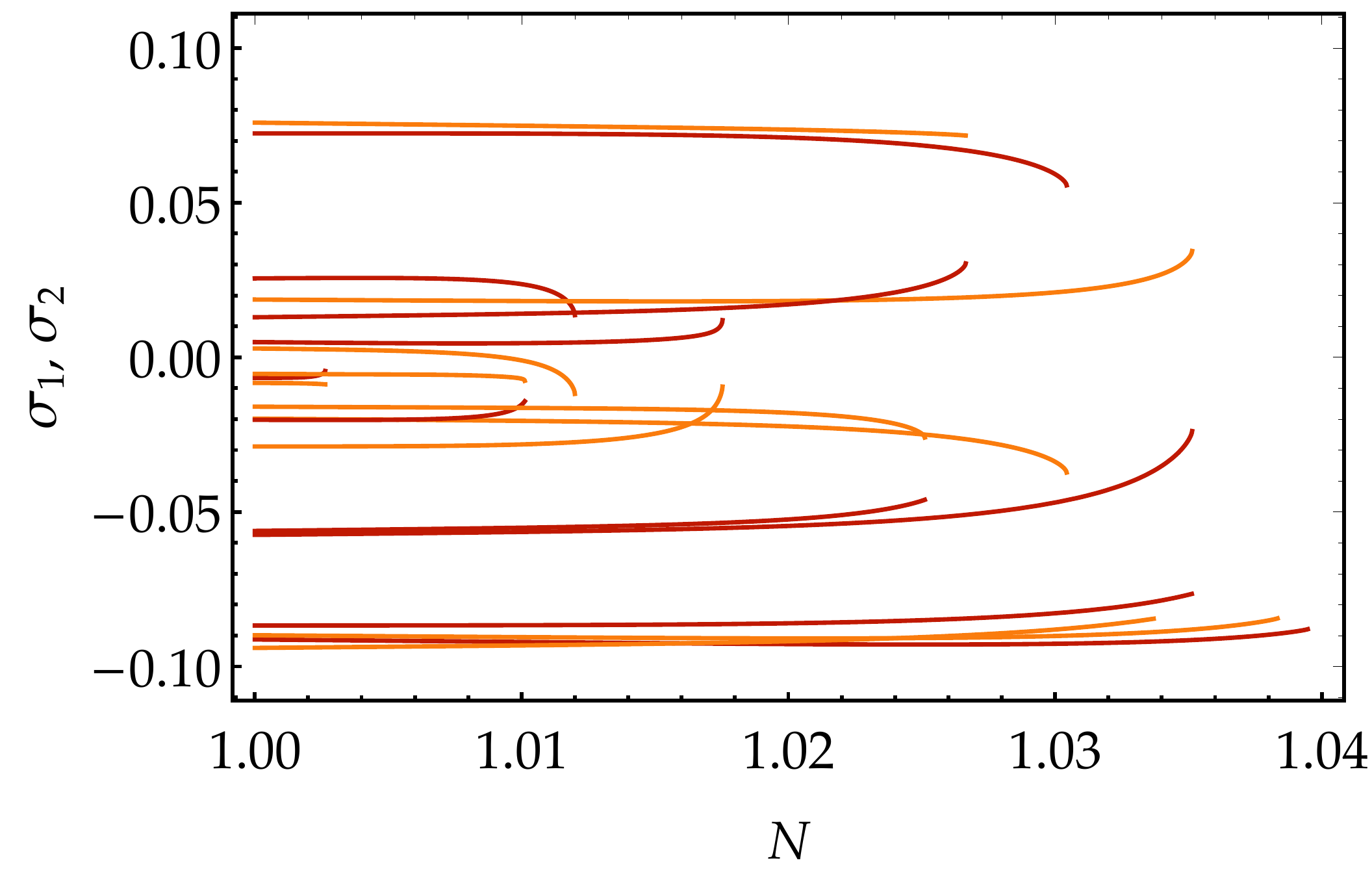}
\caption{In these plots we represent the evolution of the perturbed solution around a isotropic radiation dominated background similarly as in FIG. \ref{Fig:plotdeSitter}, but now expressing the evolution in terms of the number of e-folds $N$ instead of the cosmic time $t$. For this numerical result we have normalised to the Planck mass (i.e., we take $\mpl=1$) and we have chosen $\Lambda=0.5$, $\pECG=0.01$ and the initial value $H(N=1)=1$ (which implies $\rho_0=160.519$ due to \eqref{eq: Isotropic condition}).} \label{Fig:plotRadiation}
\end{figure}

With all of this in mind, we proceed in a similar way as in the previous section. For the numerical computation we use the initial value of the Hubble constant, $H(N_\text{ini})$, as an input and employ it to determine $\rho_0$ through the isotropic equation \eqref{eq: Isotropic condition}.\footnote{When solving for $\rho_0$ for the given value of $H(N_\text{ini})$, there are more than one branch of solutions in general. Actually, for the set of parameters employed in Fig. \ref{Fig:plotRadiation}, there is another real branch where the isotropic solution for $H$ is an increasing function. For that case, the same conclusions can be reached.}
Since we are interested in a radiation dominated era, we will neglect the cosmological constant term (initially, $\rho_0 \eN^{-4} \ll \Lambda$). In Fig. \ref{Fig:plotRadiation} we show the evolution for ten sets of randomly generated initial values for $\sigma_{1,2}$. As in the case discussed in the previous section, the numerical solutions exhibit an important deviation with respect to the isotropic background. However, now we observe no tendency in the blue curves to return to the isotropic curve. Again, we have checked that these ten curves are representative of the general behaviour. 

\section{Pathologies in higher order generalised quasi-topological theories} \label{sec: analysis quasitop}

The pathologies discussed in the previous section for the ECG case are in principle expected in any of the generalised quasitopological extensions that follow the same defining principle, i.e., the absence of certain perturbative degrees of freedom around some spacetime backgrounds. It is not difficult to convince oneself that the very same obstructions discussed above for those solutions will be a general feature of these theories. In this Section we will extend our analysis to the generalisations introduced in \cite{Arciniega:2018tnn} to explicitly show the expected problems. Interestingly, we will find that the cosmological solutions based on these extended GQTG are even more prone to problems than the ECG in a sense that we explain in the following. 

The theories that we will analyse consist in the ECG \eqref{eq: action} plus a series of higher order terms in the curvature. For our purposes here it will be sufficient to restrict our analysis to the first three higher order terms described by
\begin{equation} \label{eq: DeltaS quasitop}
  \Delta\mS=  \int \dd^4x \sqrt{|g|} \left( \frac{\beta_4}{\mpl^4} \mathcal{R}_{(4)} +\frac{\beta_5}{\mpl^6} \mathcal{R}_{(5)} +\frac{\beta_6}{\mpl^8} \mathcal{R}_{(6)} \right)\,,
\end{equation}
where $\beta_i$ ($i=4,5,6$) are dimensionless parameters and $\mathcal{R}_{(i)}$ ($i=4,5,6$) are the generalised quasi-topological curvature invariants given in \cite{Arciniega:2018tnn} and that we reproduce in Appendix \ref{appendixB} for completeness. Following the same procedure as in the ECG theory, we can obtain the evolution equations for the shear $\sigma_{1,2}$ which now have the form (compare to \eqref{eq: principal part}) 
\renewcommand\arraystretch{1.3}
\begin{align}\label{eq: pp quasitop}
\left[ \dot{H}^2 v_1 \ \mathds{1}_{2\times 2} + \eanis \left(\frac{\pECG}{\mpl^2}{\bf M}_1 +  {\bf N}_1 \right) \right] 
    \begin{pmatrix}\dddot{\sigma}_2 \\\dddot{\sigma}_1 \end{pmatrix}+
    \left[ \dot{H} v_2 \ \mathds{1}_{2\times 2}  + \eanis \left(\frac{\pECG}{\mpl^2}{\bf M}_2 +{\bf N}_2 \right) \right] 
    \begin{pmatrix}\ddot{\sigma}_2 \\\ddot{\sigma}_1 \end{pmatrix}  & \nonumber\\
    + \frac{\pECG}{\mpl^2}{\bf V} + {\bf W}
    +3 \mpl^2 \begin{pmatrix}3H\sigma_2+\dot{\sigma}_2 \\ 3H\sigma_1+\dot{\sigma}_1 \end{pmatrix} &= 0 \,.
\end{align}
\renewcommand\arraystretch{1}
where
\begin{align}
    v_1 &= -6 \frac{\beta_4}{\mpl^4} 
    - \frac{2}{5} \left(2\, H^2 + 3\, \dot{H} \right) \frac{\beta_5}{\mpl^6}
    + \frac{3}{1040} \left(16848\, H^4 + 90357\,H^2\dot{H} + 136175\, \dot{H}^2 \right) \frac{\beta_6}{\mpl^8} \,,\\
    v_2 &= -12\,  (3H\dot{H}+2\ddot{H})\frac{\beta_4}{\mpl^4}
    - \frac{4}{5} \left(6\,H^3\dot{H}+13\,H\dot{H}^2+4\,H^2\ddot{H} +9\dot{H}\ddot{H} \right) \frac{\beta_5}{\mpl^6} \nonumber \\
    & \quad + \frac{3}{520} \left(50544\, H^5\dot{H} 
    +338463\,H^3\dot{H}^2 + 589239\,H\dot{H}^3+33696\,H^4\ddot{H}+271071\,H^2\dot{H}\ddot{H}+544700\,\dot{H}^2\ddot{H}\right) \frac{\beta_6}{\mpl^8}\,,
\end{align}
and the rest of the contributions coming from $\Delta\mS$ are encoded in the matrices ${\bf N}_{1,2}$ and in the vector ${\bf W}$, which start again at zeroth order in $\eanis$.

As in the ECG case we see that the isotropic de Sitter solution ($\dot{H}=0=\eanis$) corresponds to a singular surface in phase space, thus giving rise to the same type of strong coupling problems originating from the disappearance of some dof's as discussed at length above. However, we can see that the higher order terms do not trivialise in the case of an arbitrary cosmological background ($\eanis=0$ but $\dot{H}\neq0$). This is so because the defining prescription of these theories only requires the same perturbative spectrum as GR around maximally symmetric backgrounds. As we see here however, on a general isotropic cosmological background the additional modes associated to the higher order nature of the field equations are fully active. Though this prevents any strong coupling issue, it indicates that the ghostly degrees of freedom will propagate on general cosmological backgrounds, thus making them unstable.

A possible improvement of this pathological behaviour can be obtained by noticing that there are several inequivalent terms at each order in curvature that lead to second order gravitational equations for an isotropic Ansatz, but differ beyond the isotropic solutions. In that respect, we have taken the particular combination given in \eqref{eq: DeltaS quasitop} to illustrate the present pathologies, but this is not unique nor the most general choice. Thus, it is conceivable that these terms can be combined in such a way that the equations remain of second order around arbitrary FLRW spacetimes and not only for the maximally symmetric ones, as suggested in \cite{Arciniega:2018tnn}. In other words, there could exist combinations so that the higher order contributions completely vanish in the limit $\eanis\rightarrow 0$, similarly to what happens for the ECG. This currently remains as an open question\footnote{We thank Pablo A. Cano for pointing out this possibility to us.}.

\section{Discussion} \label{sec: discussion}

In this note we have discussed the occurrence of pathologies taking place in ECG and GQTG that arise as a direct consequence of their defining prescription. These theories contain a massless spin 2, a massive spin-2 and a massive scalar fields, but are constructed so that only the two polarisations of the usual gravitational waves (the massless spin-2 field) propagate on some specific backgrounds, thus sharing the same linear spectrum as GR on those spacetimes. This condition is imposed in order to get rid of the ghostly modes associated to the higher order nature of the theories. The evanescence of degrees of freedom on these backgrounds however can be interpreted as an indication for the presence of strongly coupled modes. Since this cannot be seen at linear order, we have instead studied the full non-linear equations of slightly deformed backgrounds with fewer symmetries than the spacetimes used to define the theories. We have mainly focused on cosmological solutions and the ECG action. In this set-up, we have shown the pathological character of the isotropic solution that corresponds to a singular surface in phase space. This property generally prevents the trajectories in phase space from smoothly evolving towards the isotropic solution. Furthermore, we have discussed how standard perturbation theory around the isotropic solution fails to reproduce the full landscape of perturbations and, consequently, the conclusions drawn from a perturbative analysis cannot be fully trusted. We have also discussed these problems for the extended class of GQTG, where we have found that, not only the same strong coupling problems around the maximally symmetric backgrounds that define the theories persist, but the ghostly degrees of freedom are actually active around general cosmological backgrounds.

Although we have analysed the strong coupling problems of these theories around cosmological backgrounds, there is nothing really special about them (other than the simplicity introduced by the additional symmetries) and it is easy to envision that the same class of pathologies will be present for instance around static and spherically symmetric backgrounds. Likewise, similar problems are expected to arise in extension involving additional fields. Recently, a new class of quasi-topological electromagnetic theories has been introduced in \cite{Cano:2020qhy} where theories featuring non-minimal couplings of a $U(1)$ gauge field to gravity are explored along the lines of GQTG, i.e., with precise interactions so that the non-minimal couplings do not give rise to higher order equations of motion for a spherically symmetric Ansatz. In this respect, it is known that the so-called Horndeski vector-tensor interaction (see e.g. \cite{Horndeski:1976gi}) is the only gauge-invariant non-minimal coupling that gives rise to second order field equations (i.e. the analogue of Lovelock terms). Thus, the Lagrangians obtained in \cite{Cano:2020qhy} without additional modes on spherically symmetric backgrounds will again be prone to the same type of pathologies discussed in this work.  That would not be the case if those Lagrangians were related to the Horndeski vector-tensor interaction via a field redefinition for instance. In this respect, similar conclusions would apply to GQTG including a scalar field featuring derivative non-minimal couplings (possibly accompanied by interactions involving second order derivatives) and constructed so that the scalar only propagates one additional dof around some specific backgrounds, thus lying outside the class of Horndeski  theories or any of the known healthy scalar-tensor theories (see e.g. \cite{Zumalacarregui:2013pma,Gleyzes:2014dya,Langlois:2015cwa}).

It is interesting to notice a certain resemblance of what happens in the theories under consideration in this work and the cuscuton model, first introduced in \cite{Afshordi:2006ad} (see \cite{Iyonaga:2018vnu} for an extended version), that describes a scalar field with the peculiarity that its propagation speed becomes infinite around homogeneous configurations and the scalar field does not propagate. As in the case of the ECG and its extensions, this theory exhibits some remarkable properties. However, the fact that the field becomes non-dynamical around homogeneous backgrounds would suggest that this is a singular surface in phase space that is similar to what occurs for the defining backgrounds of GQTG. This feature was analysed in detail from a full Hamiltonian approach in \cite{Gomes:2017tzd} where it was shown that indeed the homogeneous configuration corresponds to a singular surface in phase space. It is likely that something analogous will happen for GQTG, i.e., the specific backgrounds used to the define the theories by imposing the disappearance of some modes are expected to present similar pathological properties as the homogeneous configurations in the cuscuton theory. It was argued in \cite{Gomes:2017tzd} that the cuscuton could be defined in a sensible manner only if the homogeneity of the field was imposed {\it a priori}. It would be interesting to analyse if a similar interpretation could be employed for the ECG and its generalised quasi-topological extensions. A straightforward application of the procedure would be to constrain the allowed metrics as to satisfy the desired requirements (i.e., maximally symmetric, FLRW...). Perturbations of the metric in that case would no longer be arbitrary, but they should be restricted to the corresponding subspace of metrics.

As a final remark, these theories with a reduced spectrum around some backgrounds are sometimes interpreted within the realm of Effective Field Theories (EFT). In this respect, we find it convenient to stress that, with that philosophy in mind, one should include all operators complying with the symmetries (diffeomorphims in this case) and field content (the metric and, possibly, a matter sector). In particular, there is no reason not to include the quadratic terms in the curvatures, which also introduce ghostly degrees of freedom and would become dominant at a lower scale, and higher order curvature terms that would then be order one whenever the ECG operator becomes non-perturbative. Thus, either one explicitly resorts to a non-standard and consistent organisation of the operators in the EFT or some hierarchical fine tuning is introduced. Another important question that remains is how stable the precise tuning of the coefficients in the generalised quasi-topological theories are against quantum corrections including both graviton and matter loops. It is interesting to note that all higher order curvature terms with up to two covariant derivatives acting on the Riemann can be related to the generalised quasi-topological Lagrangians via field redefinitions, as it was explicitly shown in \cite{Bueno:2019ltp}. This seems to suggest, that the generalised quasi-topological Lagrangians could serve as a basis for the gravitational EFT operators, at least partially for operators not involving higher than second derivatives of the curvatures. It would be interesting to explore how this basis would relate to the one provided in \cite{Ruhdorfer:2019qmk}.

Undoubtedly, the general class of GQTG exhibit a series of remarkable properties that make them very interesting and worth investigating. However, it is crucial to bear in mind that their very defining property is intimately related to the presence of pathologies that need to be properly tackled to guarantee the physical viability of models based on these theories such as inflationary scenarios. Of course, this should not preclude exploiting their exceptional properties to draw physically sensible and useful results from these theories. In this sense, they can be compared to other theories with a similar status. For instance, FLRW cosmological solutions\footnote{As a matter of fact, spatially flat or closed FLRW solutions do not exist in massive gravity and only open homogeneous and isotropic universes are allowed.} in massive gravity are plagued by strong coupling issues and non-linear ghost-like instabilities (see e.g. \cite{Gumrukcuoglu:2011zh,DeFelice:2012mx}). Obviously, the existence of these pathological solutions is not an obstruction for the applicability and interesting properties of massive gravity. A similar status can be granted to GQTG, although in this case the full theory does generically exhibit a pathology caused by the Ostrogradsky ghost unlike massive gravity where this problem is not present. The usefulness of the ECG and GQTG theories to obtain important physical results has actually been proven in e.g.  \cite{Bueno2019,Bueno:2020odt} where the exceptional properties of these theories allowed to obtain analytical results for the thermodynamical properties of Taub-NUT solutions that in turn helped understanding properties of general Conformal Field Theories. Thus, the main conclusion from our analysis and the pathologies that we have shown for some backgrounds simply indicate that those particular solutions cannot be regarded, at least a priori, as physical in the sense that they cannot describe real physical systems. 

~

\bf{Acknowledgments}: The authors would like to thank Pablo A. Cano and Pablo Bueno for very useful comments and feedback. JBJ acknowledges support from the  {\textit{Atracci\'on del Talento Cient\'ifico en Salamanca}} programme and the MINECO's projects PGC2018-096038-B-I00 and FIS2016-78859-P (AEI/FEDER).  This article is based upon work from COST Action CA15117, supported by COST (European Cooperation in Science and Technology). AJC is also supported by the MINECO through a PhD contract of the programme FPU 2015 with reference FPU15/02864, and the project FIS2016-78198-P.

\appendix

\section{A toy example} \label{app: toyexample}
In order to illustrate the problematic nature of dwelling on a surface where the principal part of the equations is singular, we will analyse a simple mechanical example. This will also allow us to illustrate the problem with generating perturbative solutions around a singular surface in phase space. Thus, let us consider a system with one degree of freedom $q(t)$ that evolves according to the following equation\footnote{This is not a Hamiltonian system, but this property is not relevant for our purposes here where we want to illustrate the problematic nature of solutions where the principal part vanishes.}:
 \be
 q(t) q''(t)+\Big(1-q'(t)\Big)q'(t)=0.
 \ee
 This equation has a singular surface given by $q=0$ where the principal part vanishes, which in turn is an exact solution. The general solution can be written as
 \be
 q(t)=C_1+C_2 \eN^{-t/C_1}
 \label{eq:Generalq}
 \ee
as can be checked by direct substitution. We can see that the family with $C_2=0$ reproduces the obvious constant solutions. However, among those constant solutions, the trivial one $q=0$ would require $C_1=0$, which, as we can see from the general solution, corresponds to the paradigmatic example of an essential singularity. This clearly shows that the trivial solution actually dwells on a singular surface of the space of solutions so one can expect to find difficulties to obtain perturbative solutions around it. To clearly see this, let us try to perturbative solve around $q=0$ so we expand
\be
q(t)=q^{(1)}(t)+q^{(2)}(t)+q^{(3)}(t)+\cdots
\ee
where $q^{(n)}(t)$ is assumed to be of order $n$ in some expansion parameter. It is not difficult to see that the term with second derivatives always contributes at order $(n+1)$ so that it plays no role in determining $q^{(n)}(t)$. This is analogous to what happens in the perturbative expansion around FLRW, where the terms with 4-th order derivatives never appear at $n-$th order in perturbations. This is clearly an indication that the perturbative expansion will fail in exploring the whole space of solutions around the trivial one. In our simple example, it is immediate to check that the solution for $q^{(n)}(t)$ is always a constant mode so the full perturbative solutions is
\be
q=c^{(1)}+c^{(2)}+c^{(3)}+\cdots
\ee
i.e., only the constant mode of the general solution is generated. On the other hand, if we expand around a constant but non-trivial solution $q(t)=q_0$, the perturbative solution can be obtained to be
\bea
q(t)=&&q_0+c_2^{(1)}+c_2^{(2)}+c_2^{(3)}+\cdots\nonumber\\
&+&\left[q_0 \Big(c_1^{(1)}+c_1^{(2)}+c_1^{(3)}\Big)+\Big(c_1^{(1)}c_2^{(1)}+c_2^{(1)}c_1^{(2)}+c_1^{(1)}c_2^{(2)}\Big)\left(1+\frac{t}{q_0}\right)+c_1^{(1)}\big(c_2^{(1)}\big)^2\frac{t^2}{2q_0^3}+\cdots\right]\eN^{-t/q_0}
\eea
which reproduces the expansion of the general solution \eqref{eq:Generalq} around $C_1=q_0$ and $C_2=0$, as it should. Notice that this perturbative solution is singular for $q_0=0$, thus showing once again the singular character of the trivial solution.

\begin{figure}
\includegraphics[width=0.49\linewidth]{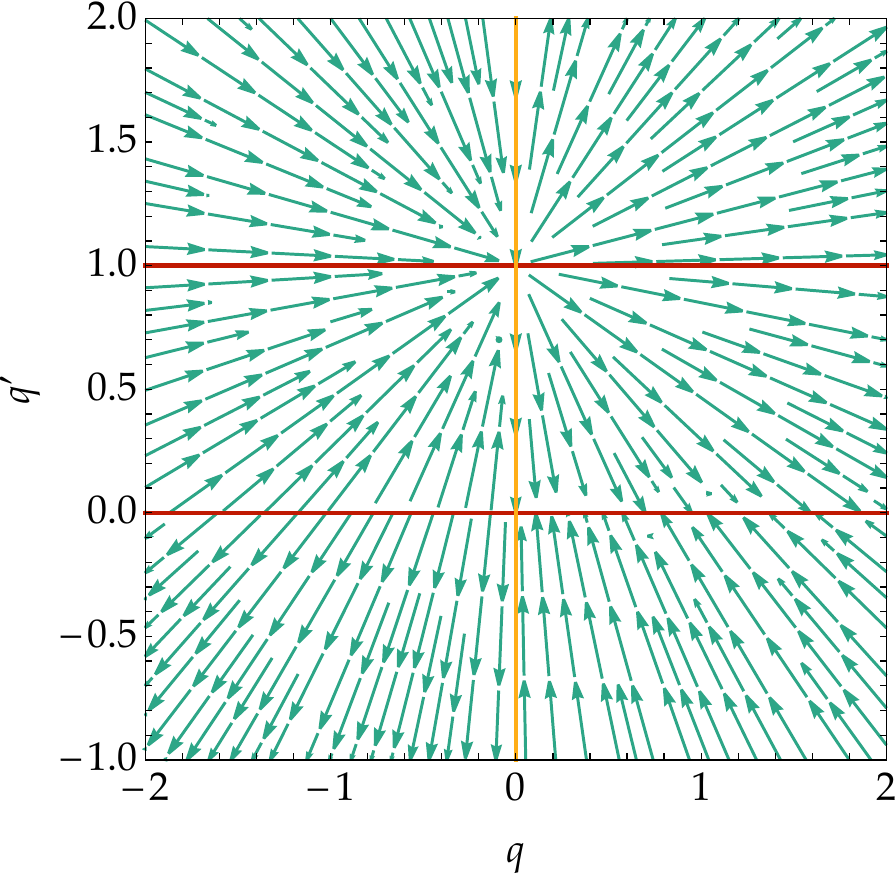}
\caption{In this Figure we show the phase map of the toy example given in \eqref{eq:Generalq}. We can clearly see how the surface $q=0$ corresponds to a separatrix in phase space as expected. }
\label{Fig:phasemapq}
\end{figure}

The second order equation  \eqref{eq:Generalq} can be written in an autonomous first order system as
\be
q'=p,\quad\quad
p'=\frac{p-1}{q}p.
\ee
Again it is obvious that $q=0$ represents a singular surface in phase space that describes a separatrix (see Fig. \ref{Fig:phasemapq}). Furthermore, there is one critical trajectory given by $p=0$. It is not difficult to see that the trajectories are straight lines of the form $p=1+cq$. From the phase map it is apparent that the separatrix is not a good physical solution for the system. Among the problems reported above, we can see that the separatrix can never be exactly reached from any point in phase space (unless it already belongs to the separatrix) and it does not correspond to an attractor region so the system will hardly evolve towards there. This example is quite trivial but it illustrates the problems with the solutions for the ECG theories and its generalisations.

\section{Curvature invariants in GQTG}\label{appendixB}
In this appendix we reproduce the terms given in \cite{Arciniega:2018tnn}
for completeness:

\begin{align}
\mathcal{R}_{(4)}&=-\frac{1}{192}\Big[5 R^{4}-60 R^{2} Q_{1}+30 R^{2} Q_{2}-160 R C_{1}+32 R C_{2}-104 R C_{3}+272 Q_{1}^{2}-256 Q_{1} Q_{2}\nonumber\\
&\qquad\qquad +336 A_{10}+48 A_{14}\Big]
\end{align}

\begin{align}
\mathcal{R}_{(5)}&=-\frac{1}{5760}\Big[15 R^{5}-36 R^{3} Q_{1}-224 R^{3} Q_{2}-336 R^{2} C_{1}-140 R^{2} C_{2}+528 R^{2} C_{3}-592 R Q_{1}^{2}+1000 R Q_{1} Q_{2}\nonumber\\
&\quad\qquad\qquad +301 R Q_{2}^{2}-912 R A_{2}-928 R A_{10}+1680 R A_{14}+1152 Q_{1} C_{1}+264 Q_{1} C_{2}+312 Q_{2} C_{2}-64 Q_{1} C_{3} \nonumber\\
&\quad\qquad\qquad -2080 Q_{2} C_{3}+4992 I_{1}\Big]
\end{align}

\begin{align}
\mathcal{R}_{(6)}& =\frac{1}{3594240}\Big[ 56813 R^{6}-523188 R^{4} Q_{1}+6234 R^{4} Q_{2}+798849 R^{3} C_{2}-558622 R^{3} C_{3}+1235848 R^{2} Q_{1}^{2} \nonumber\\
&\qquad\qquad\qquad -163250 R^{2} Q_{1} Q_{2}+42084 R^{2} Q_{2}^{2}-707808 R^{2} A_{2}+231048 R^{2} A_{10}+439920 R^{2} A_{14}-5265366 R Q_{1} C_{2} \nonumber\\
&\qquad\qquad\qquad +23208 R Q_{2} C_{2}+4902132 R Q_{1} C_{3}+44880 R Q_{2} C_{3}-704400 Q_{1}^{3}+289200 Q_{1} Q_{2}^{2}-62400 Q_{2}^{3} \nonumber\\
&\qquad\qquad\qquad +1168128 R I_{1}+792000 Q_{1} A_{2}+374400 Q_{2} A_{2}-723600 Q_{2} A_{10}-676800 C_{1}^{2}+7903368 C_{1} C_{2} \nonumber\\
&\qquad\qquad\qquad -8581680 C_{1} C_{3}-3782484 C_{2}^{2}+15454692 C_{2} C_{3}-12753720 C_{3}^{2}\Big]
\end{align}
with 
\begin{align}
&Q_1 = R_{\mu\nu}R^{\mu\nu} \,, \qquad 
 Q_2 = R_{\mu\nu\rho\sigma}R^{\mu\nu\rho\sigma} \,,\nonumber\\
&C_1 = R_\mu{}^\rho{}_\nu{}^\sigma R_\rho{}^\tau{}_\sigma{}^\eta R_\tau{}^\mu{}_\eta{}^\nu \,, \qquad
 C_2 = R_{\mu\nu}{}^{\rho\sigma} R_{\rho\sigma}{}^{\tau\eta} R_{\tau\eta}{}^{\mu\nu} \,, \qquad
 C_3 = R_{\mu\nu\rho\lambda} R^{\mu\nu\rho}{}_\sigma R^{\lambda \sigma} \,,\nonumber\\
&A_2 =R_{\mu}{}^{\sigma}{}_{\rho}{}^{\tau} R^{\mu\nu\rho\lambda} R_{\nu \alpha\lambda \beta} R_{\sigma}{}^{\alpha}{}_{\tau}{}^{\beta} \,, \quad
 A_{10} =  R^{\mu\nu} R_{\mu}{}^{\rho}{}_{\nu}{}^{\lambda} R_{\sigma\tau \alpha\rho} R^{\sigma\tau \alpha}{}_{\lambda} \, , \quad
 A_{14} =  R^{\mu\nu} R^{\rho\lambda} R_{\sigma\rho\tau\lambda} R^{\sigma}{}_{\mu}{}^{\tau}{}_{\nu} \,,\nonumber\\
&I_1 = R_{\rho\sigma}{}^{\mu\nu} R_{\mu\tau}{}^{\rho\lambda} R_{\alpha \gamma}{}^{\sigma\tau} R_{\nu\delta}{}^{\alpha \beta}R_{\lambda \beta}{}^{\gamma\delta} \,.
\end{align}

\bibliography{ECGbibliography}

\end{document}